\documentclass[%
 reprint,
superscriptaddress,
 amsmath,amssymb,
 aps,
prb,
]{revtex4-2}

\usepackage{soul}
\soulregister{\cite}7
\soulregister{\ref}7
\usepackage{color,xcolor}
\usepackage[T1]{fontenc}
\usepackage{graphicx}
\usepackage{dcolumn}
\usepackage{bm}

\usepackage[colorlinks,
linkcolor=blue,
anchorcolor=blue,
citecolor=blue,
urlcolor=blue]{hyperref}
\begin{document}
	
	\preprint{APS}
	
	\title{{An Addressable and Tunable Module for Donor-based Scalable Silicon Quantum Computing}}
	
	\author{Shihang Zhang}
    \affiliation{Shenzhen Institute for Quantum Science and Engineering, Southern University of Science and Technology, Shenzhen 518055, China}
	\affiliation{International Quantum Academy, Shenzhen 518048, China}
	\affiliation{Guangdong Provincial Key Laboratory of Quantum Science and Engineering, Southern University of Science and Technology, Shenzhen 518055, China}
	\author{Yu He}
	\email{hey6@sustech.edu.cn}
    \affiliation{Shenzhen Institute for Quantum Science and Engineering, Southern University of Science and Technology, Shenzhen 518055, China}
    \affiliation{Hefei National Laboratory, Hefei 230088, China}
	\affiliation{International Quantum Academy, Shenzhen 518048, China}
	\affiliation{Guangdong Provincial Key Laboratory of Quantum Science and Engineering, Southern University of Science and Technology, Shenzhen 518055, China}
	\author{Peihao Huang}  
	\email{huangph@sustech.edu.cn}
    \affiliation{Shenzhen Institute for Quantum Science and Engineering, Southern University of Science and Technology, Shenzhen 518055, China}
    \affiliation{Hefei National Laboratory, Hefei 230088, China}
	\affiliation{International Quantum Academy, Shenzhen 518048, China}
	\affiliation{Guangdong Provincial Key Laboratory of Quantum Science and Engineering, Southern University of Science and Technology, Shenzhen 518055, China}

	\date{\today}
	
	\begin{abstract}

 Donor-based spin qubit offers a promising silicon quantum computing route for building large-scale qubit arrays, attributed to its long coherence time and advancements in nanoscale donor placement. However, the state-of-the-art device designs face scalability challenges, notably in achieving tunable two-qubit coupling and ensuring qubit addressability. Here, we propose a surface-code-compatible architecture, where each module has both tunable two-qubit gates and addressable single-qubit gates by introducing only a single extra donor in a pair of donors. We found that to compromise between the requirement of tunability and that of addressability, an asymmetric scheme is necessary. In this scheme, the introduced extra donor is strongly tunnel-coupled to one of the donor spin qubits for addressable single-qubit operation, while being more weakly coupled to the other to ensure the turning on and off of the two-qubit operation. The fidelity of single-qubit and two-qubit gates can exceed the fault-tolerant threshold in our design. Additionally, the asymmetric scheme effectively mitigates valley oscillations, allowing for engineering precision tolerances up to a few nanometers. Thus, our proposed scheme presents a promising prototype for large-scale, fault-tolerant, donor-based spin quantum processors.
 
	\end{abstract}
	
	\maketitle
	
	\section{Introduction}\label{section:intro}
	
	Since Kane's 1998 proposal, donor-based spin qubit has gained significant attention due to its long coherence times and great potential for leveraging scalable silicon foundry manufacturing \cite{Kane1998,Morton2008,Fuechsle2010,Tyryshkin2011,Fuechsle2012,Pla2012,Saeedi2013,Muhonen2014,Harvey-Collard2017,He2019,Madzik2022}. With recent great advances of device fabrication techniques, including the ion implantation on metal-oxide-semiconductor (MOS) devices~\cite{Pacheco2017,Jamieson2017,Holmes2019,Jakob2020,Holmes2024} and scanning tunneling microscope (STM) lithographed devices~\cite{Fuechsle2012,Ward2017,WangX2020,He2019,Bussmann2021,Kiczynski2022}, the large-scale donor-based electron spin qubit platform is poised as a promising silicon quantum computing candidate~\cite{Madzik2022,Stemp2024,Thorvaldson2024}. However, achieving a scalable universal quantum processor involves more than just advanced device fabrications \cite{Loss1998,DiVincenzo2000}. Quantum systems are error-prone during computing, necessitating error correction for universal quantum computing. The most widely accepted error correction protocol is surface code, which requires a gate fidelity at the level of 99\% \cite{WangDavid2011,Fowler2012}. Several silicon-based spin qubits have already demonstrated fidelities of single-qubit and two-qubit gates exceeding this threshold, such as spin qubits in gate-defined quantum dots (QDs) and donor-based nuclear spin qubits \cite{Xue2022, Madzik2022, Noiri2022,Mills2022,HuangJY2024,WangCA2024}. Nonetheless, fault-tolerant two-qubit gates of donor-based electron spin qubits have not been realized yet. Performing high-fidelity qubit operations in a scalable quantum computer requires the tunability of the two-qubit coupling and the addressability of individual qubits.
	
	The tunability of the two-qubit coupling is particularly difficult in the donor-based system, as it requires the precision fabrication to manage exchange coupling and mitigate the valley oscillations \cite{Kalra2014,Koiller2001,Gamble2015,Salfi2018,Voisin2020,Joecker2021}. Several schemes have been proposed to improve the performance of two-qubit gates of donor-based spin qubits. For example, the electron spin qubit can be bound to a multi-donor site, which reduces the sensitivity to charge noise and decreases the valley oscillation of the exchange coupling \cite{Koiller2001,Hsueh2014,Wang2016,Buch2017,Krauth2022,Sarkar2022,Kranz2023,Jones2023}. Despite these benefits, the tunability of the two-qubit coupling is still limited in these schemes, and the short distance between the qubits is challenging even for STM lithography~\cite{WangX2020,Bussmann2021}. Alternative schemes employ an intermediate dot, an extra donor (QD) between qubits, to induce a highly tunable interaction named `superexchange' \cite{Mehl2014,Srinivasa2015,Baart2017,Rancic2017,Martins2017,Croot2018,Malinowski2018,Malinowski2019,Qiao2021,Deng2020,Chan2022,Chan2023,Munia2023}. The superexchange schemes offer advantages in reducing the gate density and the cross-talks by allowing larger spacing between the computing donors. Strong superexchange between qubits can be achieved via a donor or QD, providing enhanced flexibility in tuning the effective coupling strength~\cite{Srinivasa2015,Rancic2017,Deng2020,Chan2022,Chan2023}.
	
	The other obstacle for the scalable donor-based electron spin qubit array is achieving the addressability for each computing donor during single-qubit operations. Homogeneity makes donor-based spin qubits stable but difficult to distinguish \cite{Tyryshkin2011,Muhonen2014}. The traditional electron spin resonance (ESR) technique, which relies on the alternating magnetic field, is difficult to generate and control locally, making it challenging to address individual qubits. Utilizing the alternating electric field offers a solution by enabling more localized control via the electric dipole-induced spin resonance (EDSR) mechanism \cite{Golovach2006,Tokura2006,Tosi2017,Yoneda2018}. EDSR can be achieved through either intrinsic or artificial spin-orbit coupling (SOC). However, the requisite SOC of the electron is intrinsically weak in silicon. Furthermore, the artificial SOC is realized by engineering a more complex structure, which might also induce new noise sources \cite{Yoneda2018,Struck2020,Takeda2021}. To date, electrical control has been successfully demonstrated only for flip-flop qubits in donor systems~\cite{Reiner2024,Savytskyy2023}, while electrical control of electron spin qubits alone has not yet been achieved. Another approach to have addressable control is engineering frequency discrimination in the encoded qubits~\cite{Kane1998,Hill2015,Veldhorst2017,Hansen2022}. By tuning qubit frequencies, idling qubits can avoid unintended excitation due to off-resonance effects. The qubit frequency can be modified for the donor system using either a micromagnet or the hyperfine interaction between the electron and nuclear spins \cite{Kalra2014,Hill2015,Hile2018,Fricke2021}. For example, the qubit frequency of the nuclear spin can be distinguished by loading an electron on the donor~\cite{Hill2015,Hill2021}. Alternatively, the effective hyperfine interaction can be modified by adjusting the detuning of the electron chemical potential between two adjacent donors \cite{Tosi2017,Krauth2022}, allowing the qubit frequencies detuned accordingly.
	
	

\begin{figure}[!htbp]
		 \centering
		 \includegraphics[width=0.45\textwidth]{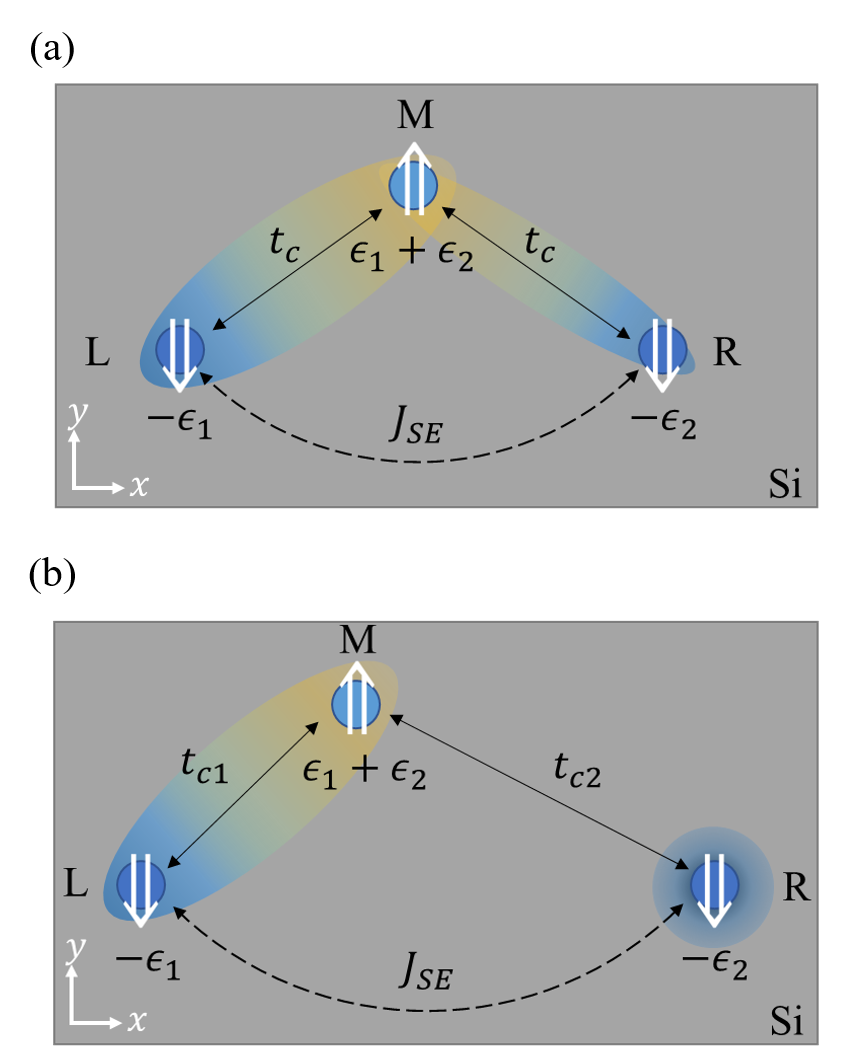}%
		 \caption{The schematic of two kinds of computing modules for donor-based spin qubits. The detuning of energies between the electron on $\mathrm{CD_{L}}$ ($\mathrm{CD_{R}}$) and $\mathrm{AD_{M}}$ is $\epsilon_{1}$ ($\epsilon_{2}$). The blue-and-yellow area represents the superposition of two single-electron charge states: one corresponding to an electron on CD and the other to an electron on AD. The blue-shaded and circular area in (b) represents another electron bounded to $\mathrm{CD_{R}}$. (a) The symmetric computing module includes two CDs and one AD. The tunneling between CDs and the AD is $t_{c}$. The AD serves as the mediator for the two-qubit coupling and provides addressability for the qubits of CDs. (b) The asymmetric computing module also includes two CDs and one AD. However, the tunneling $t_{c1}$ between $\mathrm{AD_{M}}$ and $\mathrm{CD_{L}}$ is distinct from the tunneling $t_{c2}$ between $\mathrm{AD_{M}}$ and $\mathrm{CD_{R}}$. It is assumed that $t_{c1} > t_{c2}$. In this configuration, the AD provides addressability only for the qubit of $\mathrm{CD_{L}}$.}
		 \label{fig:1}
		 \end{figure}

    In our work, we aimed to achieve a surface-code-compatible architecture, where both the tunability of two-qubit coupling and the addressability using only one ancilla donor within a two-qubit module, as shown in Fig. \ref{fig:1}(a). In the module, the donor that binds the encoding electron is called a `computing donor' (CD), while the central donor is called an `ancilla donor' (AD). The tunable superexchange coupling between spin qubits on the CDs can be induced via the AD. Additionally, the nuclear spin polarization of the CD (AD) is initialized downward (upward), allowing the qubit frequencies of the electron spins to be distinguished by adjusting the detuning between the CD and AD, similar to the method described in \cite{Tosi2017, Krauth2022}. Consequently, the addressability of the qubit can be achieved. However, we found that achieving high compatibility between the addressability and the tunability is very challenging in this computing module. A fault-tolerant single-qubit gate with addressability requires strong tunneling between the CD and AD, which prevents effectively turning off the superexchange between qubits. As a result, the residual two-qubit coupling reduces the fidelity of the single-qubit gate.
 
	Here, we propose an asymmetric scheme to solve the problem of seemingly contradictory requirements of the tunability and the addressability. For simplicity, the left (right) CD is named as $\mathrm{CD_{L}}$ ($\mathrm{CD_{R}}$), while the middle AD is named as $\mathrm{AD_{M}}$. As illustrated in Fig. \ref{fig:1}(b), $\mathrm{CD_{R}}$ is placed farther from $\mathrm{AD_{M}}$, compared with the scheme in Fig. \ref{fig:1}(a). In this scheme, the AD provides the addressability only for the qubit of $\mathrm{CD_{L}}$. Due to the weaker tunneling between $\mathrm{CD_{R}}$ and $\mathrm{AD_{M}}$, the two-qubit coupling can be effectively turned off during the single-qubit operation on the qubit of $\mathrm{CD_{L}}$. However, the two-qubit coupling can still be turned on to perform high-speed SWAP gate or controlled-Z (CZ) gate by adjusting the detunings between donors. Therefore, fault-tolerant single-qubit gates and two-qubit gates can be achieved with great compatibility between the tunability and addressability. Furthermore, the asymmetric scheme can resist the valley oscillation of the tunneling, requiring a nanoscale placement precision ($\sim5$ nm) of donors. Such placement precision can be realized by STM lithography technology, and can also be achieved when implanting heavier atoms and molecule ions using ion implantation technology \cite{Morello2020,Jakob2020,Holmes2024}. Additionally, increasing the separation between the CDs reduces the requirement on the gate density. Moreover, for the asymmetric scheme, the sweet spot still exists. Consequently, our asymmetric scheme provides a practical and highly flexible way to realize a scalable, surface-code-compatible donor-based spin qubit array for fault-tolerant quantum computing. Notably, in this design, the single ancillary donor acts as both a qubit frequency modulator and a tunable coupler, resulting in minimal resource overhead in the device fabrication process.
	
	This paper is developed as follows: In Sec. \ref{section: model}, the computing module with asymmetric structure is introduced in detail. Based on that, the addressability and the tunability of the superexchange coupling via the AD are discussed in Sec. \ref{sec: ancilla donor}. In Sec. \ref{sec:tunneling}, the flexibility of the tunneling requirement in the asymmetric scheme is illustrated, which demonstrates the tolerance for the placement precision of the donor. In Sec. \ref{sec: Architecture}, we discuss the scalable, surface-code-compatible architecture of the quantum processor based on the asymmetric computing module. In the last section, we conclude our proposal and give our perspectives on the future of the donor-based qubit processor.
	\section{Model}\label{section: model}
	
	 \begin{figure*}[!htbp]
	 \centering
	 \includegraphics[width=1\textwidth]{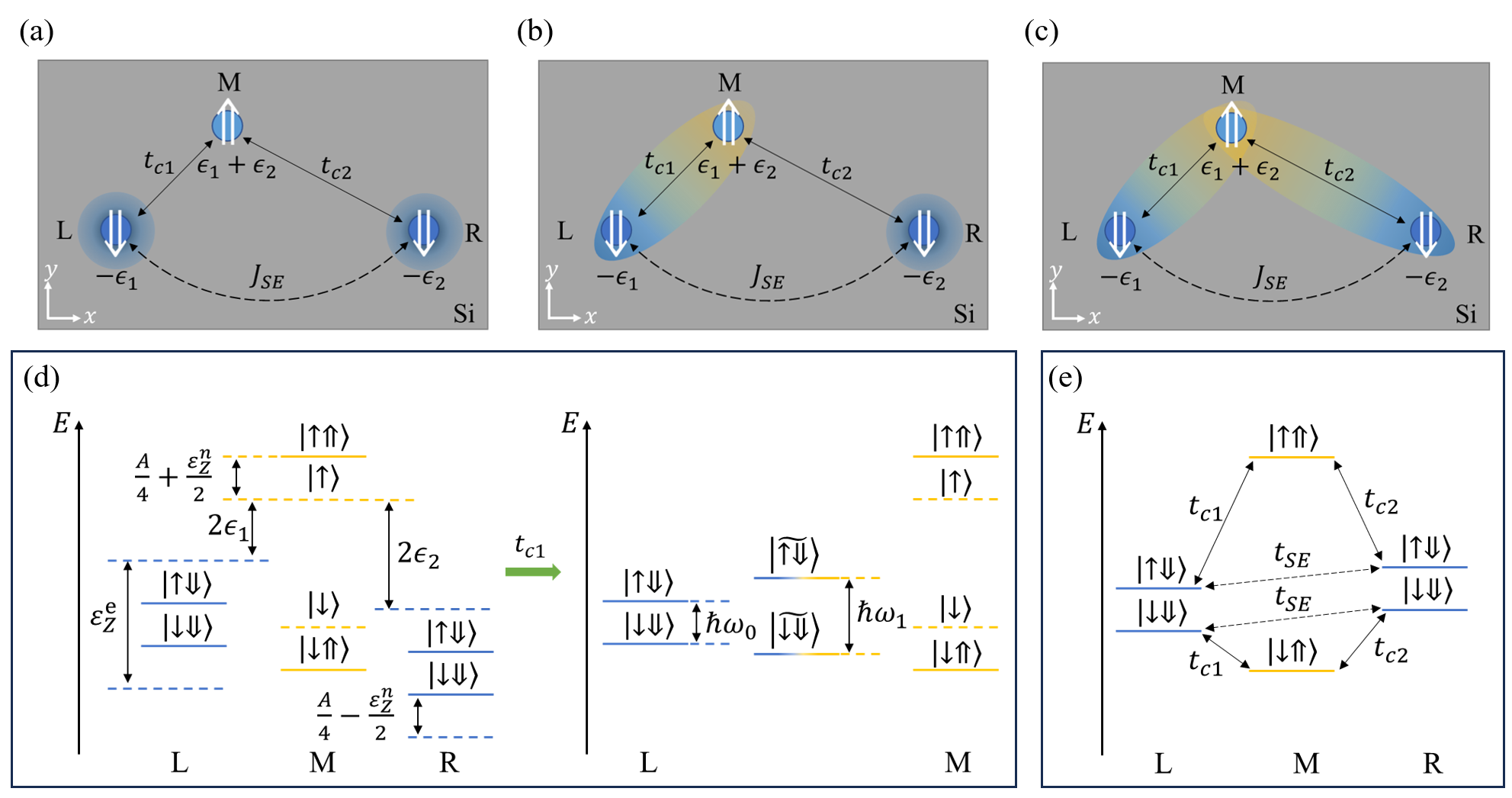}%
	 \caption{Schematic diagrams of configurations for implementing idling, addressable single-qubit operations, and tunable two-qubit operations of qubits in the asymmetric donor module. The lowest charge state of electrons is $|(1,0,1)\rangle$. For figures in the upper row, the double vertical arrow $\Uparrow$ ($\Downarrow$) represents the upward (downward) nuclear spin. (a) Computing module when both qubits are idling. Electrons are confined to their corresponding CDs. (b) Computing module when performing the addressable operation on the qubit of $\mathrm{CD_{L}}$. The left electron is in the superposition state of the $|(1,0,1)\rangle$ state and the $|(0,1,1)\rangle$ state. The right electron is confined to $\mathrm{CD_{R}}$. (c) Computing module when the two-qubit operation is executed. The system is in superposition of various two-electron charge states. (d) The shift of energy levels due to the tunneling $t_{c1}$ and detuning $\epsilon_{1}$. Dashed lines indicate energy levels without hyperfine interaction. Blue (orange) solid lines represents energy levels with downward (upward) nuclear spin. $|\tilde{\uparrow}\rangle$ ($|\tilde{\downarrow}\rangle$) represents spin state of the electron at the superposition of $|(1,0,0)\rangle$ state and $|(0,1,0)\rangle$ state. (e) Energy levels corresponding to the operational point for the two-qubit gate.}
	 \label{fig:2}
	 \end{figure*} 
	 	
	In the following section, for simplicity, a computing module is given as an example to illustrate the role of the AD. As shown in Fig. \ref{fig:1}, a computing module includes one empty AD, two CDs, and two electrons, each bound to one of the CDs. In Fig. \ref{fig:1}(b), $\mathrm{AD_{M}}$ is closer to $\mathrm{CD_{L}}$ compared to $\mathrm{CD_{R}}$. A static magnetic field is applied along the $\hat{z}$ direction (pointing out of the plane in Fig. \ref{fig:1}(b)). The chemical potential of the donor sites can be adjusted by metallic gates. The static Hamiltonian for the computing module is as follows: 
	\begin{equation}\label{Eq:H}
	H = H_{e}+H_{n}+H_{hf},
	\end{equation} 
	where $H_{e}$ is the Hamiltonian describing the orbital and spin part of electrons, $H_{n}$ is the Zeeman Hamiltonian for the nuclear spin, and $H_{hf}$ is the Hamiltonian describing the hyperfine interaction between the electron spin and nuclear spin. 
 
    For two electrons in a three-donor system, the Hamiltonian $H_{e}$ is modeled by the generalized Fermi-Hubbard Hamiltonian:
    \begin{equation}
    \begin{aligned}
    H_{e} = &\frac{\gamma_{e} B_{0}}{2} \sum_{i} (\hat{n}_{i,\uparrow} - \hat{n}_{i,\downarrow}) - \sum_{\langle i,j \rangle,\sigma} t_{i,j} (\hat{c}_{i,\sigma}^{\dagger} \hat{c}_{j,\sigma} + \hat{c}_{j,\sigma}^{\dagger} \hat{c}_{i,\sigma}) \\
             & +\sum_{i,\sigma} \mu_{i} \hat{n}_{i,\sigma} 
             + U \sum_{i} \hat{n}_{i,\downarrow} \hat{n}_{i,\uparrow} 
             + V \sum_{\langle i,j \rangle} \hat{n}_{i} \hat{n}_{j},
    \end{aligned}
    \end{equation}
	where $i=L,M,R$ indicates the electron is bound to $\mathrm{CD_{L}}$, $\mathrm{AD_{M}}$, and $\mathrm{CD_{R}}$. $\sigma=\uparrow,\downarrow$ indicates the spin polarization of electrons in the $\hat{z}$ direction. The first term is the Zeeman Hamiltonian of electrons. $\hat{n}_{i,\sigma}$ is the number operator for electrons. $\hat{c}_{i,\sigma}^{\dagger}$ and $\hat{c}_{i,\sigma}$ are the creation and annihilation operators of electrons, respectively. $\mu_{i}$ is the electrochemical potential of electron on site $i$. $t_{i,j}$ is the tunneling of electron between site $i$ and site $j$. In this work, we assume that detunings $\epsilon_{1}=2(\mu_{M}-\mu_{L})$, $\epsilon_{2}=2(\mu_{M}-\mu_{R})$, and tunnelings $t_{c1}=t_{L,M}$, $t_{c2}=t_{M,R}$. $U$ is the on-site Coulomb repulsion between electrons \cite{Fuechsle2010}. $V$ is the nearest neighbor Coulomb repulsion between electrons \cite{Kiczynski2022}. $\gamma_{e}$ is the electron spin gyromagnetic ratio. $B_{0}$ is the strength of the static magnetic field. 

    $H_{n}=\frac{\gamma_{n}B_{0}}{2}\sum_{i}(\hat{n}_{i,\Uparrow}-\hat{n}_{i,\Downarrow})$ is the Zeeman Hamiltonian for the nuclear spin. $\hat{n}_{i,\Uparrow/\Downarrow}$ is the number operator of nucleus with spin in the $\hat{z}$ direction on site $i$. $\gamma_{n}$ is the nuclear spin gyromagnetic ratio. The spin eigenstates of the Hamiltonian $H_{n}$ are $\mid\Downarrow\rangle$ and $\mid\Uparrow\rangle$. The nuclear spin state of the CD (AD) is initialized to be $\mid\Downarrow\rangle$ ($\mid\Uparrow\rangle$).

    The hyperfine interaction between nuclear spins and electron spins can be described by $H_{hf}$. A more general form of the hyperfine Hamiltonian is listed in Sec. I of the supplementary materials~\cite{supp}. In the following discussion, a simplified version of $H_{hf}$ is considered:
    \begin{equation}\label{Eq:H-hf}
	H_{hf} = A[(\hat{n}_{M,\uparrow}-\hat{n}_{M,\downarrow})\hat{n}_{M,\Uparrow}+\sum_{i}^{L,R}(\hat{n}_{i,\downarrow}-\hat{n}_{i,\uparrow})\hat{n}_{i,\Downarrow}],
	\end{equation}
	where $A$ is the strength of the hyperfine interaction between the electron spin and nuclear spin for single P donor in silicon. According to the equation, when the spin polarizations of the electron and nucleus are aligned (opposite), the energies of electron spin states become higher (lower). This is key to ensure the addressability of qubits.
    
    In Eq. \eqref{Eq:H-hf}, only ZZ-coupling of $H_{hf}$ are considered in the HF Hamiltonian. This is justified since the coherence time of the nuclear spin is significantly longer than that of the electron spin. Moreover, flip-flops between electron-nuclear spin states can be suppressed due to the off-resonance between flip-flop qubits and computing qubits \cite{Tosi2017,Savytskyy2023,Reiner2024}. The frequency detunings between flip-flop qubits and the computing qubit are approximately tens of MHz, depending on the difference between hyperfine interaction strength and nuclear spin Zeeman splitting. Additionally, the drive amplitude of the flip-flop process can be reduced by selecting an appropriate driving field orientation. Then, the flip-flop term in the hyperfine interaction can be neglected, resulting in the ZZ coupling form in Eq. \eqref{Eq:H-hf}.
	
	The basis states of the total Hamiltonian can be constructed by the spin states and orbital states of electrons. The orbital states $|(n_{L},n_{M},n_{R})\rangle$ of electrons include single-occupied and double-occupied states of two electrons at three sites. $n_{L}$ ($n_{M}$, $n_{R}$) is the number of electrons on $\mathrm{CD_{L}}$ ($\mathrm{AD_{M}}$, $\mathrm{CD_{R}}$). The spin states consist of two spin-1/2 states, defined as qubits in this work. For example, $|(\uparrow,0,\downarrow)\rangle$ represents one electron with spin-up at $\mathrm{CD_{L}}$ and one electron with spin-down at $\mathrm{CD_{L}}$. In this work, electrons are primarily in the $|(1,0,1)\rangle$ state, which is set as the lowest charge state. However, due to the tunneling between donors, when the detuning between donors is adjusted, the wavefunction of the electron bounded to donors is distorted. For example, if the detuning $\epsilon_{1}$ between $\mathrm{CD_{L}}$ and $\mathrm{AD_{M}}$ is $-\frac{V}{2}$, the charge state becomes an equal superposition of the $|(1,0,1)\rangle$ state and the $|(0,1,1)\rangle$ state. In this work, the qubit frequency of the left electron in the $|(1,0,1)\rangle$ state is defined as $\hbar\omega_{0}$, while the modified qubit frequency of the electron in superposition states is defined as $\hbar\omega_{1}$. These charge states play an important role in achieving the scalability of the computing module.

	\section{Asymmetric donor module} \label{sec: ancilla donor}
	In this work, the asymmetric donor module is proposed for scalable qubit devices. As mentioned in the introduction, both addressability and tunability can be achieved with ADs. The most natural and direct idea is the symmetric scheme, as shown in Fig. \ref{fig:1}(a). However, we found a significant contradiction between the addressability and the tunability in the symmetric scheme. To solve the problem, the AD provides the addressability only to the its nearest qubit in our proposed asymmetric scheme.

	In this section, for simplicity, only one electron and two donors are considered for discussion on the addressability. Since the nuclear spin state of CD (AD) is initialized to be $\mid\Downarrow\rangle$ ($\mid\Uparrow\rangle$), the state of the system including one single electron can be represented by a combination of electron spin states and nuclear spin states. For example, the state $\mid\updownarrow\Downarrow\rangle_{\mathrm{L}}$ ($\mid\updownarrow\Uparrow\rangle_{\mathrm{M}}$) indicates that the electron is bound to $\mathrm{CD_{L}}$ ($\mathrm{AD_{M}}$), where the $\mid\updownarrow\rangle$ represents the electron spin being in the spin-up or spin-down state, and $\mid\Downarrow\rangle_{\mathrm{L}}$ ($\mid\Uparrow\rangle_{\mathrm{M}}$) represents the spin polarization of nucleus at $\mathrm{CD_{L}}$ ($\mathrm{AD_{M}}$). Moreover, when the two-electron states is in the superposition of $|(\updownarrow,0,\updownarrow)\rangle$ state and $|(0,\updownarrow,\updownarrow)\rangle$ state, the state of the left electron and the corresponding donors can be simplified to $|\widetilde{\updownarrow\Downarrow}\rangle$, which is superposition of  $\mid\updownarrow\Downarrow\rangle_{\mathrm{L}}$ state and $\mid\updownarrow\Uparrow\rangle_{\mathrm{M}}$ state.
	
	The manipulation schemes and corresponding energy levels of the asymmetric computing module are illustrated in Fig. \ref{fig:2}. In Fig. \ref{fig:2}(a), both qubits are idling with electrons bound to the CDs. In Fig. \ref{fig:2}(b), by adjusting the detuning $\epsilon_{1}$ between $\mathrm{CD_{L}}$ and $\mathrm{AD_{M}}$, the distribution of the electron spreads to $\mathrm{AD_{M}}$. Consequently, as shown in Fig. \ref{fig:2}(d), energies of electron spin states are modified. The modified qubit frequency $\hbar\omega_{1}$ can be utilized to provide addressability for single-qubit operations. When two-qubit operations are required, as shown in Fig. \ref{fig:2}(c), wavefunctions of the two electrons spread to $\mathrm{AD_{M}}$ to induce effective tunneling. The effective tunneling $t_{SE}$ can induce the so-called superexchange by adjusting detunings \cite{Baart2017}. Based on the superexchange, two-qubit operations can be achieved. In the following, we show how addressability and tunable superexchange can be realized by introducing a single AD in a computing module. 

	In this section, we introduce and discuss the asymmetric donor modules with ADs. In Sec. \ref{section:3A}, the addressability of the qubits is discussed based on a two-donor system. Moreover, the contradiction between the addressability and the tunability in the symmetric scheme is analyzed also in Sec. \ref{section:3A}. In Sec. \ref{section:3B}, the tunability of the two-qubit coupling in the asymmetric structure is discussed. In Sec. \ref{section:joff}, we show how the contradiction between the tunability and addressability can be avoided in the asymmetric scheme.
	
	For the following discussion, we assume: $B_{0}=0.3$ T, $\gamma_{e}=27.97$ GHz/T, $\gamma_{n}=-17.41$ MHz/T, and $A\approx 117$ MHz for single P donor in silicon; the on-site Coulomb repulsion $U\approx 43.6$ meV \cite{Fuechsle2010}, the nearest off-site Coulomb repulsion $V\approx 4$ meV \cite{Kiczynski2022}. Realistic noise parameters are considered: charge noise with $\sigma_{\epsilon}=2$ $\mu$eV \cite{Jock2018,Kranz2020} and a hyperfine-induced dephasing rate of 1 kHz. In this work, we set $0.5\%$ as the error threshold for quantum gates. In Sec. \ref{sec: conclusion}, the possibility of achieving a fidelity above $99.9\%$ for single-qubit gates is further discussed.
	
	 \begin{figure*}[!htbp]
	 \centering
	 \includegraphics[width=1\textwidth]{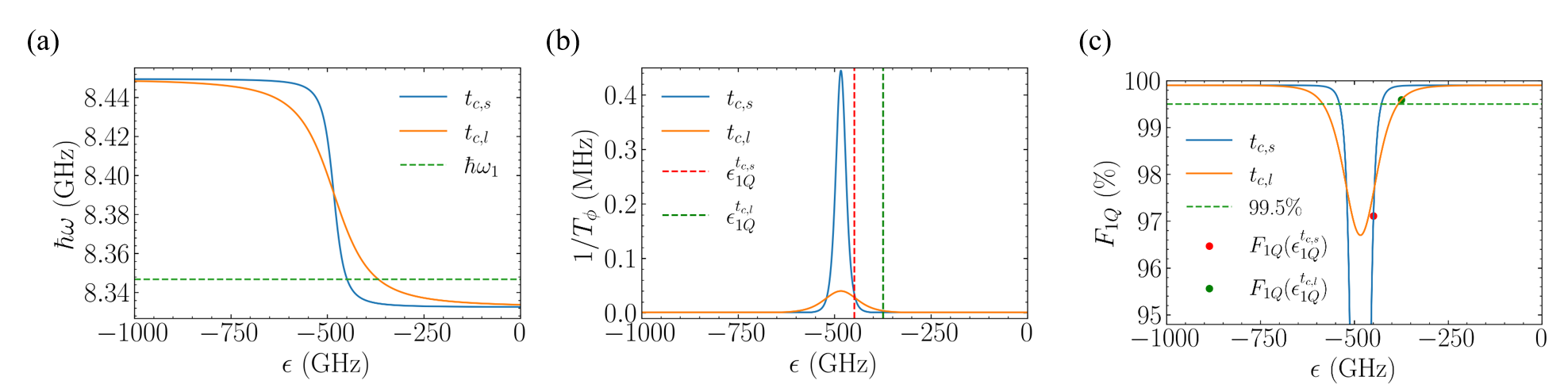}%
	 \caption{The addressability of computing qubits for small tunneling $t_{c,s}=30$ GHz (blue lines) and large tunneling $t_{c,l}=100$ GHz (orange lines). (a) Qubit frequency $\hbar\omega$ is plotted as a function of the detuning $\epsilon$ between the CD and the AD. The green dashed line indicates the qubit frequency with the cross-talk error of $0.5\%$, where the qubit is operated at the corresponding detuning. (b) The dephasing rate $\frac{1}{T_{\phi}}$ due to the charge noise is plotted as a function of the detuning $\epsilon$. The red (green) dashed line indicates the detuning where the qubit frequency is $\omega_{1}$ for the small tunneling $t_{c,s}$ (the large tunneling $t_{c,l}$). (c) Fidelity of the single-qubit gate $F_{\mathrm{1Q}}$ is plotted as a function of the detuning $\epsilon$. The red (green) point indicates the computing point for the small tunneling $t_{c,s}$ (the large tunneling $t_{c,l}$). The fidelity at the point can exceed $99.5\%$ (green dashed line) for the larger tunneling of $t_{c,l}$.}
	 \label{fig:3}
	 \end{figure*} 
	
	\subsection{Addressability during single-qubit operations}\label{section:3A}
	The addressability of computing qubits in our scheme is achieved through qubit frequency discrimination. The modification of the computing qubit frequency is based on the hyperfine interaction between the electron spin and nuclear spin and detunings between donors. For simplicity, in this subsection, only $\mathrm{CD_{L}}$ and $\mathrm{AD_{M}}$ are considered. As mentioned in the previous section, the hyperfine interaction increases (decreases) the energies of the electron spin states with parallel (anti-parallel) nuclear spin polarization. For instance, when the nuclear spin polarization of the CD is initialized downward, loading an electron onto the CD results in a decrease (increase) in the energy of the spin-up (spin-down) state of the electron. Consequently, the qubit frequency of the electron spin is decreased, compared to that without the hyperfine interaction. For the same reason, the qubit frequency of the electron bound to the AD is increased. By adjusting the detunings between electrochemical potentials of electrons on the donors, the wavefunction of the electron can redistribute between the CD to the AD. In other words, electrons transit from the $|(1,0,1)\rangle$ state to the superposition of the $|(1,0,1)\rangle$ state and the $|(0,1,1)\rangle$ state. The spin state corresponding to the superposition state is defined as $|\widetilde{\updownarrow\Downarrow}\rangle$. The frequency of the qubit on the $\mid\updownarrow\Downarrow\rangle_{\mathrm{L}}$ states is defined as idling qubit frequency $\hbar\omega_{0}$. As shown in Fig. \ref{fig:2}(d), the modified qubit frequency $\hbar\omega_{1}$ is the energy difference between the states $|\widetilde{\uparrow\Downarrow}\rangle$ and $|\widetilde{\downarrow\Downarrow}\rangle$ due to the hyperfine interaction. The qubit frequency undergoes modifications as the detuning is adjusted ($\hbar\omega_{1}>\hbar\omega_{0}$). By applying an ESR pulse that resonates with the modified qubit frequency, specific qubits in a qubit array are operated. Due to the off-resonance between the qubit frequency $\hbar\omega_{0}$ and the driving frequency $\hbar\omega_{1}$, operations on idling qubits can be suppressed, as shown in Sec. V of the supplementary materials~\cite{supp}. Hence, the addressability for the computing qubits could be achieved. The fidelity of addressable single-qubit gates and idling gates will be evaluated in Sec. \ref{section:joff}.
	
	We assume the detuning between the chemical potential of the electron on the CD and AD is $\epsilon$. The relationship between the qubit frequency $\hbar\omega$ and the detuning $\epsilon$ can be obtained by numerical diagonalization of the Hamiltonian. As shown in Fig. \ref{fig:3}(a), the qubit frequency is plotted as a function of the detuning $\epsilon$. We assume that the tunneling between the CD and the AD is $t_{c,s}=30$ GHz and $t_{c,l}=100$ GHz. The qubit frequency undergoes modification within a range governed by the hyperfine interaction, approximately $A\approx 117$ MHz. To prevent unwanted coupling between qubits, the electron is not completely transferred to the AD. In this section, we assume a single-qubit gate time of 1 $\mu$s. To effectively reduce the error induced by unwanted driving on idling qubits to a mere $0.5\%$, the detuning between frequencies of computing qubits is assumed to be around $10\sqrt{2}$ ($\approx14$) MHz for addressing individual qubits (see Sec. V of the supplementary materials~\cite{supp}).
	
	A noticeable issue is that the qubit frequency becomes sensitive to charge noise since the state of the electron is a superposition of charge states on the two donors, as illustrated in Fig. \ref{fig:2}(d). When the wavefunction of the electron is tuned between donors, the inclusion of charge-excited states may induce errors due to leakage or decoherence. Note that under the magnetic control, qubit states would not leak to the charge-excited state easily. However, as shown in Fig. \ref{fig:3}(b), the qubit dephasing rate $\frac{1}{T_{\phi}}$ increases as the detuning is adjusted to the charge transition point between the $|(1,0,1)\rangle$ state and the $|(0,1,1)\rangle$ state ($\epsilon \approx -480$ GHz). During the single-qubit operation, charge noise induces the qubit dephasing, which can potentially dominate qubit decoherence. As a result, the fidelity of the single-qubit gate can be reduced. As shown in Fig. \ref{fig:3}(c), the fidelity of the single-qubit gate does not exceed the fault-tolerant threshold at the working point with the required detuning for the addressability with small tunneling $t_{c,s}$ (Detailed estimations of the error are provided in Sec. IV of the supplementary materials~\cite{supp}). To suppress the effect of charge noise, a larger tunneling is necessary to ensure both addressability and long-lived coherence of the electron spin qubits. For example, as shown in Fig. \ref{fig:3}(a), the qubit frequency increases more smoothly as the detuning increases with a larger tunneling $t_{c,l}$. As shown in Fig. \ref{fig:3}(b), the dephasing rate of the qubit $\frac{1}{T_{\phi}}$ becomes smaller. As a result, the fidelity of the single-qubit gate exceeds $99.5\%$ for $t_{c,l}$, as shown in Fig. \ref{fig:3}(c). Thus, a large tunneling of the electron between the CD and AD ensures the fault-tolerant single-qubit gate, while preserving the addressability of individual qubits.		
	
	\textit{Contradiction in the symmetric scheme.} Another role of the AD is to serve as a mediator for the `superexchange' between qubits. Before discussing the superexchange in the asymmetric scheme, we illustrate the contradiction between the addressability and the tunability in the symmetric scheme. Superexchange is achieved through a virtual tunneling between the CDs, as shown in Fig. \ref{fig:2}(b). The high tunability of the superexchange is based on two adjustable detuning between CDs and the AD \cite{Srinivasa2015,Rancic2017,Chan2023}. However, during the single-qubit operation on one of the qubits, the corresponding detuning is fixed to enable the addressability. Consequently, the tunability of the two-qubit coupling is severely limited. Moreover, a large tunneling is required to achieve high-fidelity single-qubit operations by mitigating the effect of the charge noise. As a result, the two-qubit coupling cannot be effectively turned off in the symmetric scheme ($J\propto t_{c}^{4}$ \cite{Srinivasa2015,Rancic2017,Chan2023}). Whereas, weak tunneling is required to turn off the residual coupling. Therefore, the different requirements for tunneling are contradictory for the addressability and the tunability. We propose an asymmetric scheme to resolve the problem. In the scheme, one AD provides addressability only for one CD. The requirement on tunneling between AD and other CDs can be relaxed. With lower tunnelings between AD and other CDs, the superexchange can be turned off, which will be discussed in Sec. \ref{section:joff}. Moreover, the `sweet line' ($\epsilon_{1}=\epsilon_{2}$) suppressing the effect of charge noise on superexchange can still be utilized for the CZ gate, in our scheme, where the qubit dephasing due to the variation on the detuning $\epsilon_{1}-\epsilon_{2}$ is weak. The sweet line is discussed in Sec. III of the supplementary materials~\cite{supp}.

	\subsection{Tunability of the two-qubit coupling}\label{section:3B}
	
		 \begin{figure*}[!htbp]
		 \centering
		 \includegraphics[width=1\textwidth]{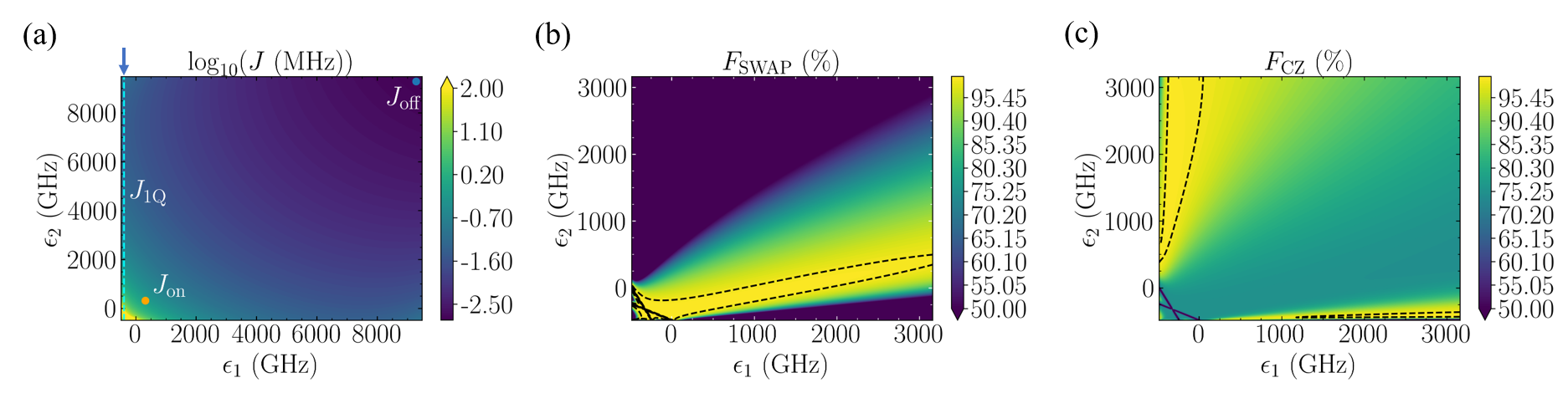}%
		 \caption{Two-qubit coupling and gate fidelities for the asymmetric scheme, when $t_{c1}=80$ GHz and $t_{c2}=20$ GHz. (a) The strength of the superexchange coupling $\log_{10}(J)$ is plotted as a function of detunings $\epsilon_{1}$ and $\epsilon_{2}$. The blue point ($J_{\mathrm{off}}$) represents the operational point for idling gates of both qubits. The orange point ($J_{\mathrm{on}}$) represents the operational point for two-qubit operations, while the cyan dashed line represents the operational line for single-qubit operations of qubit 1 (indicated by the blue arrow). (b) Fidelity of the SWAP gate as a function of detunings $\epsilon_{1}$ and $\epsilon_{2}$. The high-fidelity area ($F_{\mathrm{SWAP}}>99\%$) is enclosed by black dashed lines. High-fidelity SWAP gates can be obtained around the region with small detuning between qubit frequencies when $(\epsilon_{1}-\epsilon_{0})/(\epsilon_{2}-\epsilon_{0})\approx t_{c1}/t_{c2}$. (c) Fidelity of the CZ gate as a function of detunings $\epsilon_{1}$ and $\epsilon_{2}$. The high-fidelity area ($F_{\mathrm{CZ}}>99\%$) is enclosed by black dashed lines. High fidelity can be obtained around the region with strong superexchange coupling and large detuning between qubit frequencies when the condition $(\epsilon_{1}-\epsilon_{0})/(\epsilon_{2}-\epsilon_{0})< t_{c1}/t_{c2}$ is satisfied.}
		 \label{fig:5}
		 \end{figure*}
		 
	We have illustrated the necessity of the asymmetric scheme. In this subsection, we discuss the tunability of the two-qubit coupling in the asymmetric scheme, decoherence, cross-talk errors, and the corresponding gate fidelities. By adjusting detunings between the CDs and the AD, superexchange between spin qubits can be induced and tuned. Using the forth-order Schrieffer-Wolff (SW) transformation, the coupling between the $|(\downarrow,0,\uparrow)\rangle$ state and the $|(\uparrow,0,\downarrow)\rangle$ state can be obtained. The effective exchange coupling (i.e. superexchange) is (see Sec. III of the supplementary materials~\cite{supp})
	\begin{equation}
	J_{\mathrm{SE}} = t_{c1}^{2}t_{c2}^{2}\beta(\epsilon_{1},\epsilon_{2}),
	\end{equation}
	where $t_{c1}$ ($t_{c2}$) are the tunneling between the $\mathrm{CD_{L}}$ ($\mathrm{CD_{R}}$) and $\mathrm{AD_{M}}$. The expression $J_{\mathrm{SE}}\propto t_{c1}^{2}t_{c2}^{2}$ indicates that the process of the superexchange involves forth-fold virtual tunneling events. The parameter $\beta(\epsilon_{1},\epsilon_{2})$ depends on the detunings $\epsilon_{1}$ and $\epsilon_{2}$. $\beta$ represents differences between energy levels within tunneling events (see Sec. III of the supplementary materials for details~\cite{supp}). In general, $\beta$ increases as detunings $\epsilon_{1}$ or $\epsilon_{2}$ decreases towards the charge transition point between the $|(1,0,1)\rangle$ state and $|(1,1,0)\rangle$ state or $|(0,1,1)\rangle$ state.
	
	To identify the optimal operation points for P-donor-based electron spin qubits in silicon, we also estimate the fidelity of two-qubit gates, considering the effects of charge noise, nuclear spin noise, and detuning between the two qubits. For charge-noise-induced dephasing, the variation of the detuning between donors is assumed as $\sigma_{\epsilon}=2$ $\mu$eV~\cite{Jock2018,Kranz2020} (see Sec. IV of the supplementary materials~\cite{supp} for calculation of the dephasing rate). Additionally, the dephasing rate due to the nuclear spin noise is assumed as 1 kHz \cite{Tyryshkin2011}. In the presence of the finite two-qubit coupling, the suppression of the SWAP operation between spin states due to the frequency detuning between qubits are discussed in Sec. V of the supplementary materials~\cite{supp}. The strength of the superexchange coupling can also be obtained by numerical diagonalization. The analytical and numerical results are in good agreement except around charge transition lines between $|(1,0,1)\rangle$ state and other states, as shown in Sec. II of the supplementary materials~\cite{supp}.
    
    Fig. \ref{fig:5}(a) shows that the superexchange between the spin qubits is highly tunable by the detunings $\epsilon_{1}$ and $\epsilon_{2}$ (further details on the tunability can be found in Sec. II of the supplementary materials~\cite{supp}). For large positive detunings, electrons are mostly in the $|(1,0,1)\rangle$ state, where the superexchange is relatively small. As the detunings decrease towards the charge transition point, the effective exchange coupling between the qubits is enhanced, enabling the operation of two-qubit gates. 
	
    As shown in Fig. \ref{fig:5}(b) and \ref{fig:5}(c), high-fidelity SWAP gate and CZ gate can be achieved, for $t_{c1}=80$ GHz, $t_{c2}=20$ GHz. In these figures, there are different regions with the fidelity exceeding $99.5\%$. Several processes reduce the fidelity outside these regions. Firstly, for detunings close to the charge transition line ($\epsilon_{1(2)}\approx -\frac{V}{2}$), charge noise-induced dephasing becomes dominant in qubit decoherence. Secondly, for large detunings between the CDs and the AD, the two-qubit coupling is significantly suppressed. As a result, with $\epsilon_{1},\epsilon_{2}>3000$ GHz, the gate time of the SWAP gate approaches the qubit coherence time and reduces gate fidelities. The regions of the high-fidelity SWAP gate are indicated by the black dashed lines in Fig. \ref{fig:5}(b). Since the fidelity of the SWAP gate is reduced by the detuning between qubit frequencies, the high-fidelity region appears where the resonance condition $\hbar\omega_{1}=\hbar\omega_{2}$ is satisfied. The detunings $\epsilon_{1(2)}$ required for the resonance of qubits satisfy the condition $(\epsilon_{1}-\epsilon_{0})/(\epsilon_{2}-\epsilon_{0})=t_{c1}/t_{c2}$, where $\epsilon_{0}$ is the detuning at the charge transition point between CDs and the AD (see Sec. VII of the supplementary materials~\cite{supp}). Note that due to the asymmetric structure, the resonant condition for qubit frequencies is no longer satisfied at the symmetric point ($\epsilon_{1}=\epsilon_{2}$). 
    
    Although the fidelity of the SWAP gate is reduced by the detuning between qubit frequencies, the fidelity of the CZ gate is not suppressed. Fig. \ref{fig:5}(c) shows the fidelity of the CZ gate as a function of detunings $\epsilon_{1}$ and $\epsilon_{2}$. The fidelity is significantly reduced around the resonant condition $(\epsilon_{1}-\epsilon_{0})/(\epsilon_{2}-\epsilon_{0})=t_{c1}/t_{c2}$. This is because, in this case, the frequency detuning between qubits is much smaller than the superexchange coupling, which induces the unwanted rotations between spin state $\mid\uparrow\downarrow\rangle$ and $\mid\downarrow\uparrow\rangle$ in the presence of the finite superexchange interaction between the spin qubits. When the difference $\epsilon_{2}-\epsilon_{1}$ between the detunings becomes larger, the frequency detuning between qubits is larger, and in the meanwhile, the superexchange coupling becomes stronger. Hence, $F_{\mathrm{CZ}}$ is enhanced due to the strong superexchange instead of being reduced as the case of $F_{\mathrm{SWAP}}$. Fig. \ref{fig:5}(b) and (c) show that, with proper choice of parameters, the fidelities of both the SWAP and the CZ gates can be over the quantum error-correction threshold (above 99\%). Therefore, based on the asymmetric structure, the computing module is scalable with great addressability of qubits and tunability of the two-qubit coupling.
	
	\begin{figure}[!htbp]
	\centering
	\includegraphics[width=0.45\textwidth]{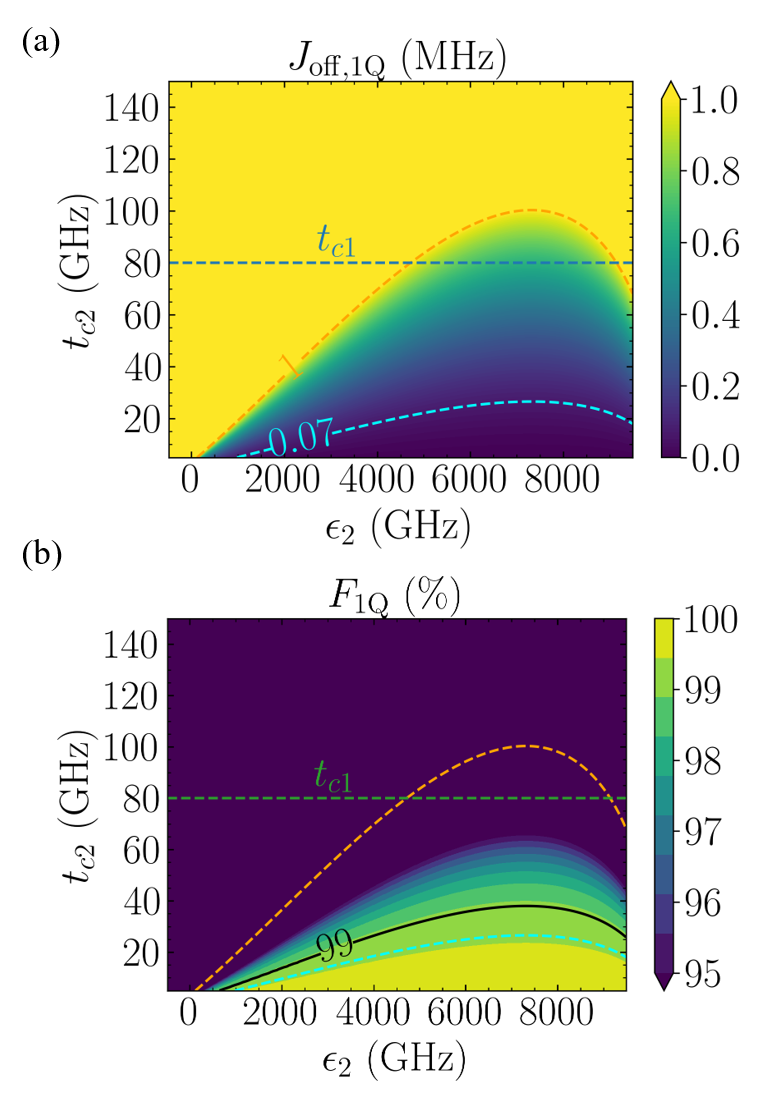}%
	\caption{The tunability of the superexchange in the asymmetric scheme. (a) The strength of the superexchange coupling during the single-qubit operation $J_{\mathrm{off,1Q}}$ is plotted as a function of $\epsilon_{2}$ and $t_{c2}$. (b) The fidelity of the single-qubit operation $F_{\mathrm{1Q}}$ is plotted as a function of $\epsilon_{2}$ and $t_{c2}$. The black solid line represents the fault-tolerant threshold ($F_{\mathrm{1Q}}=99\%$). In the figure, the detuning $\epsilon_{1}$ is set to the operation point for the left computing qubit. The blue (green) dashed line in (a(b)) corresponds to the tunneling $t_{c2}=t_{c1}=80$ GHz, which corresponds to the symmetric scheme. The orange (cyan) dashed line indicates the upper bound for reducing the cross-talk error lower than 0.5\% from the unwanted-rotation (-phase-accumulation) process.}
	\label{fig:4}
	\end{figure} 		
	
	\subsection{Turning-off two-qubit coupling during single-qubit gates}\label{section:joff}

	A critical aspect of the two-qubit coupling is its effective on-and-off ratio. As mentioned above, there is a contradiction between the addressability of qubits and the tunability of the two-qubit coupling in the symmetric scheme. Turning off the two-qubit coupling requires small tunnelings, while the addressability of the qubits necessitates large tunnelings. In the asymmetric scheme, the tunneling $t_{c1}$ between $\mathrm{AD_{M}}$ and $\mathrm{CD_{L}}$ is large, while the tunneling $t_{c2}$ is set to be smaller. Since the strength of the superexchange $J_{\mathrm{SE}}\propto t_{c1}^{2}t_{c2}^{2}$, then the two-qubit coupling can be effectively turned off when the addressed qubit is being operated.
	
	The residual superexchange coupling during single-qubit operations might induce two kinds of errors: unwanted rotations between anti-parallel spins and unwanted phase accumulation between spin-parallel states and spin-anti-parallel states (see Sec. V of the supplementary materials~\cite{supp}). During single-qubit operation, there is a detuning of $10\sqrt{2}$ MHz between qubit frequencies. With this detuning, the error of unwanted rotations can be lower than $0.5\%$ with $J_{\mathrm{off,1Q}} < 1$ MHz (see Sec. V of the supplementary materials~\cite{supp}). In this case, the unwanted phase accumulation still occurs. The superexchange coupling should be low enough that the accumulated phase is small during single-qubit operations. Considering a gate time of $1$ $\mu$s, the unwanted phase-accumulation error can be lower than $0.5\%$, with $J_{\mathrm{off,1Q}} < 0.07$ MHz (see Sec. V of the supplementary materials~\cite{supp}). Therefore, $0.07$ MHz is the threshold for high-fidelity single-qubit qubit gates. The threshold determines the requirement for turning-off the superexchange coupling. Alternatively, the threshold $J_{\mathrm{off,1Q}}$ can in principle be higher since the phase accumulation error can be mitigated by proper optimal control methods~\cite{Vandersypen2005,Khaneja2005,WangX2014,ZengJK2019,YangCH2019,Xue2022,Heinz2024}. In this work, we choose the more strict requirement ($J_{\mathrm{off,1Q}} < 0.07$ MHz) without considering optimal control methods.
	
	The flexible setting of $t_{c2}$ in the asymmetric scheme is the key to mitigating the residual two-qubit coupling during single-qubit operations. To demonstrate the importance of the asymmetric scheme, in Fig. \ref{fig:4}, $J_{\mathrm{off,1Q}}$ and $F_{\mathrm{1Q}}$ are plotted as functions of the detuning $\epsilon_{2}$ and the tunneling $t_{c2}$. We assume the detuning $\epsilon_{1}=\epsilon_{\mathrm{1Q}}$ and the tunneling $t_{c1}=80$ GHz, where the error of the addressable single-qubit gate is lower than $0.5\%$ (discussed in Sec. \ref{sec:tunneling} later). In Fig. \ref{fig:4}(b), the situation of the symmetric scheme ($t_{c2}=t_{c1}=80$ GHz) is indicated by the dashed green line, where the fidelity of the single-qubit operation is much lower than the fault-tolerant threshold. In the asymmetric scheme, the flexible setting of $t_{c2}$ can suppress the $J_{\mathrm{off,1Q}}$ to be lower than 0.07 MHz. Consequently, the fidelity is higher with smaller $t_{c2}$ in the asymmetric scheme, and $F_{\mathrm{1Q}}$ can exceed the fault-tolerant threshold. Thus, the contradiction between the addressability and the tunable two-qubit coupling can be solved in the asymmetric scheme.
	
	\section{Tolerance for the tunneling between donors} \label{sec:tunneling}
	
	 \begin{figure*}[!htbp]
	 \centering
	 \includegraphics[width=1\textwidth]{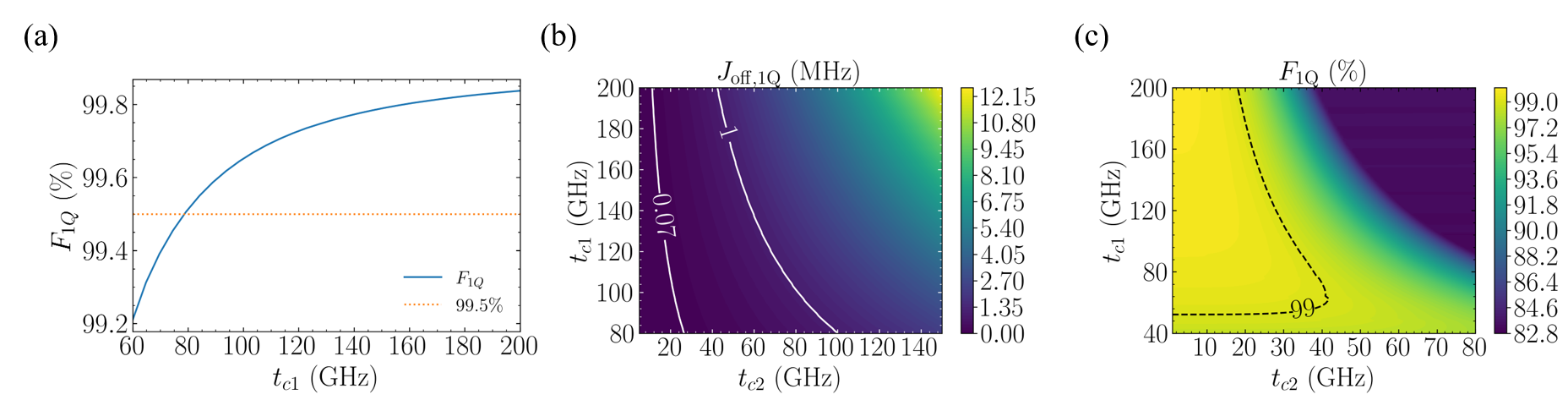}%
	 \caption{The tolerance for the tunneling in the asymmetric scheme. (a) The fidelity of single-qubit gates $F_{\mathrm{1Q}}$ is plotted as a function of the tunneling $t_{c}$ without residual exchange coupling. For consistency, the threshold for fidelity is set to $99.5\%$ (the orange dotted line). (b) $J_{\mathrm{off,1Q}}$ is plotted as a function of $t_{c1}$ and $t_{c2}$. Two white lines indicate the thresholds, determining the maximum of $t_{c2}$. The white line `1 (0.07)' indicates the upper bound of the superexchange coupling limited by unwanted two-qubit rotation (phase accumulation). (c) The fidelity of single-qubit gates on two qubits is plotted as a function of $t_{c1}$ and $t_{c2}$. The black dashed line indicates the fault-tolerant fidelity ($>99\%$). The tolerance for $t_{c2}$ depends on the strength of $t_{c1}$.}
	 \label{fig:6}
	 \end{figure*} 
		 
	It is evident from the previous section that the tunnelings between donors significantly influence the scalability and operational performance of the computing module. However, in experiments, setting and adjusting the tunneling is challenging. The tunneling exhibits oscillation with the distance between them due to valley states, especially at short distances ($<10$ nm) \cite{Koiller2001,Gamble2015,Salfi2018,Voisin2020,Joecker2021}. Even with the nanoscale placement precision achieved through STM lithography, the oscillation in the tunneling is inevitable \cite{Voisin2020}. As a result, the tolerance for the tunneling determines the feasibility of the donor-based computing module. In this section, high tolerance for the tunneling in the asymmetric scheme is demonstrated for scalable quantum computation.
	
	As mentioned previously, there are two critical issues for the scalable computing module with an AD. Firstly, the qubit becomes sensitive to the charge noise during operations. To mitigate the sensitivity, a large tunneling is required for fault-tolerant single-qubit gates. In Fig. \ref{fig:6}(a), the fidelity $F_{1Q}$ of the single-qubit gate at the working point is plotted as a function of the tunneling. In the figure, only a single qubit is considered, where cross-talk error does not contribute to the gate fidelity. Note that in a multi-qubit system, there is a cross-talk error of 0.5\% at the working point (this error is not present in Fig. \ref{fig:6}(a), but it is considered in Fig. \ref{fig:6}(c).) As shown in the figure, the fidelity $F_{1Q}$ increases as the tunneling increases. The requirement on tunneling $t_{c1}$ for the high-fidelity ($>99.5\%$) single-qubit gate is determined ($t_{c1}>80$ GHz). Secondly, turning off the two-qubit coupling is challenging around the computing point for single-qubit gates. The detuning $\epsilon_{1}$ between $\mathrm{CD_{L}}$ and $\mathrm{AD_{M}}$ is small at the computing point of single-qubit operation. This makes it difficult to turn off the two-qubit coupling because the superexchange can only be tuned by detuning $\epsilon_{2}$. To solve the issue, the tunneling $t_{c2}$ should be reduced by placing $\mathrm{CD_{R}}$ farther from $\mathrm{AD_{M}}$. In the following discussion, the upper bound to the cross-talk error is set as $0.5\%$. With a gate time of $1$ $\mu$s, the superexchange should be lower than 0.07 MHz. Then, the error corresponding to the phase accumulated during a $\pi$-rotation of a single qubit is lower than $0.5\%$ (see Sec. V of the supplementary materials~\cite{supp}). To clarify the requirement on $t_{c2}$, $J_{\mathrm{off,1Q}}$ is plotted as a function of $t_{c1}$ and $t_{c2}$ in Fig. \ref{fig:6}(b). A white line in the figure corresponds to $J_{\mathrm{off,1Q}}$ equaling 0.07 MHz, where the corresponding $t_{c2}$ is the upper bound for $t_{c2}$. The lower bound for $t_{c2}$ is determined by the fidelity of the two-qubit gate. Due to the large $t_{c1}$, the superexchange can be turned on over a wide range of $t_{c2}$. Thus, the lower bound for $t_{c2}$ is small and the corresponding boundary line is invisible in the figure.
	
	Based on the discussion in the previous paragraph, fidelity of gates $X\otimes I$ on two qubits is a key performance indicator for fault-tolerant quantum computing in the donor qubit module. As shown in Fig. \ref{fig:6}(c), the fidelity of a single-qubit gate on one of the qubits in a donor qubit computing module (including two CDs and one AD) is plotted as a function of the tunnelings $t_{c1}$ and $t_{c2}$. An $X$ gate is applied to one of the qubits, while an $I$ gate is applied to the other one. The errors in the calculation include errors from the charge noise, dephasing from surrounding nuclear spin, and cross-talk errors. Cross-talk errors could be from the residual superexchange coupling between qubits and off-resonant driving on the idling qubit. The discussion of errors is detailed in Sec. VIII of the supplementary materials~\cite{supp}. The region with high fidelity is similar with the region with the residual superexchange coupling lower than 0.07 MHz, as shown in Fig. \ref{fig:6} (c). Near the upper bound for tunneling $t_{c2}$, the unwanted phase accumulation is the dominant error. The upper bound of $t_{c2}$ is lower for larger $t_{c1}$. For example, the upper bound to the $t_{c2}$ is around 40 (30) GHz with $t_{c1} = 60$ $(100)$ GHz. While for a weaker $t_{c2}$, the unwanted phased accumulation and the unwanted operation due to the residual exchange coupling are suppressed. As a result, the dominant error is the dephasing due to charge noise or noise from the surrounding nuclear spins.
	
	With the improved tolerance for tunnelings, the requirement on the donor placement precision can be relaxed. In the asymmetric computing module, $t_{c1}$ has a higher threshold compared to $t_{c2}$. Consequently, the tolerance for variation on $t_{c2}$ (defined as $t_{c2}^{\max}/t_{c2}^{\min}$) is higher than that on $t_{c1}$. Arranging the donors in the appropriate directions can increase the tolerance to imprecision in inter-donor spacing. For example, according to Ref. \cite{Voisin2020}, the valley oscillation of the tunneling is weak along the [110] direction in silicon. Due to the lattice symmetry, the valley oscillation of the tunneling is also weak along the $[\bar{1}10]$ direction. Therefore, relative to the ancilla donors, the computing donors on either side can be planted along the [110] direction and $[\bar{1}10]$ direction. According to the theoretical results of tunneling magnitude from effective mass theory, the distance between $\mathrm{CD_{L}}$ ($\mathrm{CD_{R}}$) and $\mathrm{AD_{M}}$ can be around 10-15 (16-21) nm \cite{Joecker2021}. While referring to the theoretical results of tunneling magnitude based on tight-bind modeling, the distance can be around 7-12 (13-20) nm \cite{Wang2016,Tankasala2022}. Within these ranges of distances, the oscillation on tunneling is within one order of magnitude \cite{Voisin2020}. Consequently, the requirements for tunnelings $t_{c1}$ and $t_{c2}$ can be satisfied with state-of-the-art nanoscale donor placement technology \cite{Voisin2020,Joecker2021,Tankasala2022}. Besides, the AD requires no readout unit, which reduces the complexities in the design and fabrication of devices for large-scale quantum computing.

	\section{Quantum processor based on the asymmetric scheme} \label{sec: Architecture}
	\begin{figure*}[!htbp]
		 \centering
		 \includegraphics[width=0.9\textwidth]{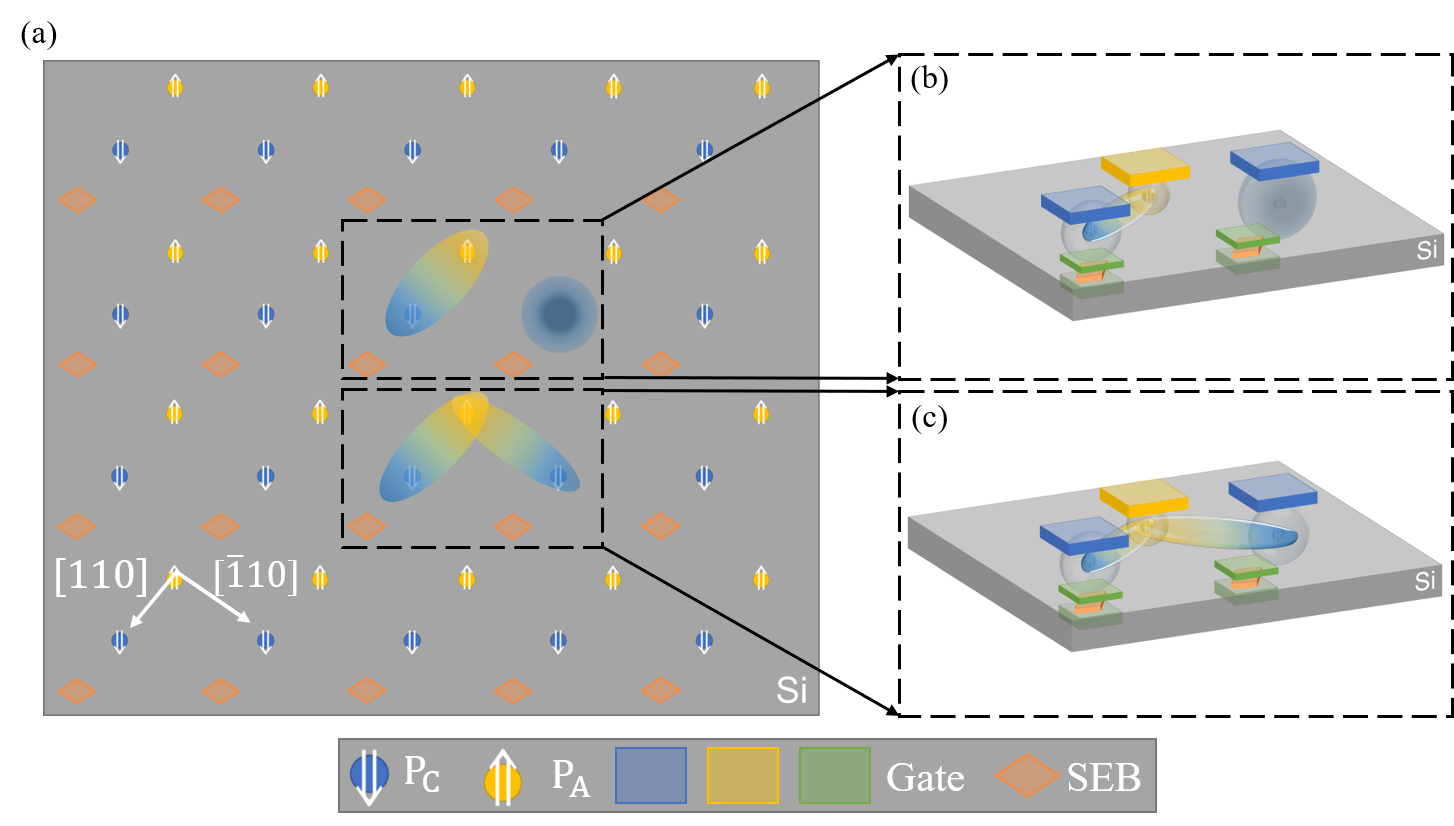}%
		 \caption{The schematic of the scalable donor quantum computing device based on the asymmetric scheme. (a) 2D donor array based on the asymmetric computing module. Ideally, for any AD, the nearest CD is positioned along the [110]-axis, while the second nearest CD is positioned along the $[\bar{1}10]$-axis. (b) 3D structure of the devices in the upper framed area in (a) for the illustration of single-qubit gate operation. Gates above atoms are designed to adjust detunings between donors. ESR and NMR pulses are applied by an antenna. Two-qubit gates can be operated between electron spin qubits bound to any atom and its adjacent atoms. Electrons are loaded to CDs by single electron boxes (SEBs) around them. One of the qubits is being operated, while the other one is idling. (c) 3D structure of the scalable qubit devices in the lower framed area in (a), where a two-qubit gate between qubits is being executed.}
		 \label{fig:7}
		 \end{figure*} 
		
	In this section, based on the asymmetric scheme, we propose a surface-code-compatible quantum processor for donor-based electron spin qubits. For gate-defined QD qubits and donor-based nuclear spin qubits, architectures for surface-code-compatible quantum processor have been proposed \cite{Hollenberg2006,Schenkel2011,Hill2021,Tosi2017,Gorman2016,Veldhorst2017,Li2018,Boter2022,Unseld2023}. The implementation of the surface code error correction requires each qubit to be connected to its nearest neighbor qubits in a two-dimensional arrangement~\cite{Fowler2012}. The architecture of the device based on the asymmetric scheme is shown in Fig. \ref{fig:7}(a). Donors are positioned in silicon as required by the asymmetric scheme. ADs (yellow) are placed line by line. Then, for each AD, two CDs (blue) are placed around ADs with different distances and orthogonal axis. The nuclear spin polarizations on CDs (ADs) are initialized downward (upward). Each AD provides addressability to the nearest CD and serves as a mediator to induce superexchange between the corresponding CD and its adjacent CD. Single-electron boxes (SEBs) are placed around CDs for initialization and measurement of qubits. Top gates positioned above donors can adjust the detunings of electrons between donors. The 3D schematic of the device is shown in Fig. \ref{fig:7}. Fig. \ref{fig:7}(b) illustrates the distributions of electrons for single-qubit operations, whereas Fig. \ref{fig:7}(c) shows the corresponding distributions for two-qubit operations. To achieve addressable single-qubit operations, the computing qubits are tuned into resonance with the frequency of ESR pulses, while idling qubits are adequately detuned to prevent unwanted driving. As shown in Fig. \ref{fig:7}(b), the electron wavefunction of the addressed qubit redistributes to the AD, while the electron of the idling qubit is tightly bound to the CD. Then, the ESR and nuclear magnetic resonance (NMR) pulses are applied to the device globally by an antenna (not shown in the figure). To suppress the unwanted flip-flop process, the electric-field component of the pulses should be aligned along the [110] direction. Moreover, the frequency detuning between the flip-flop qubit and the computing qubit also inhibit the flip-flop process. When a two-qubit operation is required, wavefunctions of both electrons redistribute to activate the two-qubit coupling by adjusting detunings, as shown in Fig. \ref{fig:7}(c). In conclusion, a scalable surface-code-compatible quantum processor can be constructed based on the asymmetric scheme.
	
	\section{Discussion and Conclusion} \label{sec: conclusion}
	To solve the contradiction between the addressability and the tunability, an asymmetric computing module of donors is proposed under state-of-art experimental conditions. It is worth noting that there are two additional options to construct scalable donor-based spin qubit devices, which are not discussed in this article. The first alternative approach is to introduce magnetic field gradient using a micromagnet. With a sufficient magnetic field gradient, qubits can be addressed by different qubit frequencies. In this scheme, the AD serves solely as a mediator of the superexchange between qubits, reducing the contradiction between the addressability and the tunability. By estimation, a magnetic gradient $b\approx0.08$ mT/nm is sufficient to construct computing modules with errors of 0.1\%. The required magnetic gradient is feasible in experiment \cite{Yoneda2015,Philips2022}. However, there are still limitations in the scheme. The frequency detuning between qubits in the asymmetric computing module can be adjusted electrically, whereas the micromagnet-induced frequency detuning between qubits is not tunable. Moreover, micromagnets might introduce extra dephasing processes and require careful design of micromagnets for field gradient uniformity and proper field strength \cite{Yoneda2018,Struck2020,Takeda2021}. As the device scaling up, the compatibility between the qubit array and the design of the micromagnet becoming more challenging. For large-scale arrays, the micromagnet scheme still faces issues with frequency crowding and related cross-talks. Also, the micromagnet is incompatible with the global control scheme~\cite{Hansen2022}, since the magnetic field gradient generated by the micromagnet can result in non-uniformity of the global field. However, the micromagnet can be compatible with the asymmetric computing module. 
	
	The second alternative approach is to introduce two more ADs into one computing module. Each additional donor is placed in close proximity to a CD and has a nuclear spin polarization opposite to that of CDs. Before single-qubit operations, the electrons of the computing qubits are transported to the extra donors, allowing the qubits to be addressed and operated. The superexchange could still be induced by the AD. This approach guarantees the addressability and the tunability as well, and avoids the contradiction between them. However, two-dimensional expansion of the module is more complex. In the approach, more donors are required in a computing module, increasing difficulties in device fabrication. Moreover, the direct tunneling of the electron between donors might induce extra decoherence~\cite{Tosi2017,Krauth2022}: fast diabatic transfer can excite the charge states, while slow transfer can cause the dephasing of qubits as the electron spin becomes sensitive to charge noise. In summary, both of these approaches are feasible for constructing a scalable quantum processor of spin qubits, but need to be further studied and improved. They are worthwhile to study in future work.
	
	In this work, the error of $0.5\%$ represents a compromise between cross-talk errors and operation errors. For example, to have lower cross-talk errors, a large detuning between qubit frequencies is required. This brings the computing point of the single-qubit operation closer to the charge transition point, leading to increased sensitivity to charge noise. Consequently, the fidelity of the single-qubit gate is reduced. There are several potential approaches to enhance the gate fidelity. The upper bound of the gate fidelity might be increased by reducing the charge noise to $\sigma_{\epsilon}\approx 0.2$ $\mu$eV \cite{Kranz2020} ($\sigma_{\epsilon}=2$ $\mu$eV in this work). Moreover, as shown in Fig. \ref{fig:6}(b), if errors on qubit operations can be corrected with residual exchange coupling (even only phase accumulation corrected), the restriction on the tunnelings can be relaxed. Considering charge noise with $\sigma_{\epsilon}=0.2$ $\mu$eV \cite{Kranz2020} and a hyperfine-induced dephasing rate of 0.5 kHz, the fidelity of single-qubit gates can exceed $99.9\%$ with even higher tolerance for $t_{c1}$. Moreover, optimizing control methods can also further enhance the fidelities of qubit gates in this setup~\cite{Vandersypen2005,Khaneja2005,WangX2014,ZengJK2019,YangCH2019,Xue2022,Heinz2024}.

	In this article, we have proposed a scalable, surface-code-compatible computing module for donor-based electron spin qubits with an asymmetric structure. This novel scheme enables addressable and fault-tolerant single-qubit gates of a specific qubit through the aid of an extra donor. The ancillary empty donor can also serve as a mediator for the coupling between qubits. The ancillary dot-enabled superexchange coupling is highly tunable and allows the coupling to be effectively turned off. The tunability of the two-qubit coupling ensures effective on-off control of the two-qubit operations. Notably, our asymmetric scheme resolves the contradiction between the addressability and the tunability of two-qubit coupling. In particular, we show that the fidelity of single-qubit and two-qubit gates can exceed the fault-tolerant threshold. Moreover, the tolerance for tunnelings between donors, based on the tunability, relaxes the precision requirement for donor placement to $\sim$5 nm. The mediator-based coupling between qubits facilitates large distances between CDs, thereby reducing manufacturing difficulties. With the ability to implement surface code error correction, this proposed scheme presents a promising prototype for a large-scale, fault-tolerant spin-based quantum processor.

	\begin{acknowledgments}
		This work was supported by the National Natural Science Foundation of China (Grants No. 92165210, 11904157, 62174076), the Innovation Program for Quantum Science and Technology (No. 2021ZD0302300), the Science, Technology and Innovation Commission of Shenzhen Municipality (Grants No. KQTD20200820113010023).
	\end{acknowledgments}
 
\bibliography{reff}
	
\end{document}